\documentclass[a4paper]{report}
\usepackage[utf8]{inputenc}
\usepackage[T1]{fontenc}
\usepackage{RJournal}
\usepackage{amsmath,amssymb,array}
\usepackage{booktabs}
\usepackage{comment}

\RequirePackage{subfigure}
\usepackage{framed}
\RequirePackage[OT1]{fontenc}
\RequirePackage{amsthm,amsmath,amsfonts,booktabs}
\RequirePackage{bm,dsfont}
\usepackage{threeparttable}
\usepackage{comment}
\usepackage[T1]{fontenc}
\usepackage{changepage}
\usepackage[utf8]{inputenc}
\usepackage{pdflscape}
\usepackage{afterpage}
\numberwithin{equation}{section}
\theoremstyle{plain} 
\theoremstyle{plain} 
\theoremstyle{plain} 
\theoremstyle{plain} 
\theoremstyle{plain} 

\theoremstyle{plain}

\theoremstyle{plain} 

\theoremstyle{plain}

\newcommand{\bX}{\bm{X}}

\newcommand{\ba}{\bm{a}}

\newcommand{\bx}{\bm{x}}
\newcommand{\bbeta}{\bm{\beta}}

\newcommand{\btheta}{\bm{\theta}}
\newcommand{\balpha}{\bm{\alpha}}

\DeclareMathOperator\bE{\mathbb E} % expectation
\DeclareMathOperator\bV{\mathbb V} % variance

\newcommand{\RR}{\mbox{\tiny{RR}}}
\newcommand{\RD}{\mbox{\tiny{RD}}}
\newcommand{\OR}{\mbox{\tiny{OR}}}
\newcommand{\aug}{\mbox{\tiny{aug}}}

\newcommand{\sumi}{\sum_{i=1}^N}

\newcommand{\sumk}{\sum_{k=1}^J}

\newcommand{\be}{\begin{eqnarray}}
\newcommand{\ee}{\end{eqnarray}}
\newcommand{\bee}{\begin{eqnarray*}}
\newcommand{\eee}{\end{eqnarray*}}
\newcommand{\bi}{\begin{enumerate}[(i)]}
\newcommand{\ei}{\end{enumerate}}

\begin{document}

%% do not edit, for illustration only
\sectionhead{Contributed research article}
\volume{XX}
\volnumber{YY}
\year{20ZZ}
\month{AAAA}

%% replace RJtemplate with your article
\begin{article}
  % !TeX root = RJwrapper.tex
\title{\pkg{PSweight}: An R Package for Propensity Score Weighting Analysis}
\author{by Tianhui Zhou, Guangyu Tong, Fan Li, Laine E. Thomas and Fan Li}

\maketitle

\abstract{
Propensity score weighting is an important tool for comparative effectiveness research. Besides the inverse probability of treatment weights (IPW), recent development has introduced a general class of balancing weights, corresponding to alternative target populations and estimands. In particular, the overlap weights (OW) lead to optimal covariate balance and estimation efficiency, and a target population of scientific and policy interest. We develop the R package \CRANpkg{PSweight} to provide a comprehensive design and analysis platform for causal inference based on propensity score weighting. \CRANpkg{PSweight} supports (i) a variety of balancing weights, (ii) binary and multiple treatments, (iii) simple and augmented weighting estimators, (iv) nuisance-adjusted sandwich variances, and (v) ratio estimands. \CRANpkg{PSweight} also provides diagnostic tables and graphs for covariate balance assessment. We demonstrate the functionality of the package using a data example from the National Child Development Survey (NCDS), where we evaluate the causal effect of educational attainment on income.
}

\section[Introduction]{Introduction} \label{sec:intro}

Propensity score is one of the most widely used causal inference methods for observational studies \citep{Rosenbaum1983}. Propensity score methods include weighting, matching, stratification, regression, and mixed methods such as the augmented weighting estimators. The \CRANpkg{PSweight} package provides a design and analysis pipeline for causal inference with propensity score weighting \citep{Robins1994, Hirano2003, Lunceford2004, LiMorganZaslavsky2018}. There are a number of existing R packages on propensity score weighting (see Table \ref{tb:summary}). Comparing to those, \CRANpkg{PSweight} offers three major advantages: it incorporates (i) visualization and diagnostic tools of checking covariate overlap and balance, (ii) a general class of balancing weights, including overlap weights and inverse probability of treatment weights, and  (iii) multiple treatments. More importantly, \CRANpkg{PSweight} comprises a wide range of functionalities, whereas each of the competing packages only supports a 
subset of these functionalities. 
As such, \CRANpkg{PSweight} is currently the most comprehensive platform for causal inference with propensity score weighting, offering analysts a one-stop shop for the design and analysis. Table \ref{tb:summary} summarizes the key functionalities of \CRANpkg{PSweight} in comparison to related existing R packages. We elaborate the main features of \CRANpkg{PSweight} below.

\CRANpkg{PSweight} facilitates better practices in the design stage of observational studies, an aspect that has not been sufficiently emphasized in related packages. Specifically, we provide a design module that facilitates visualizing overlap (also known as the positivity assumption) and evaluating covariate balance without access to the final outcome \citep{austin2015moving}. When there is limited overlap, \CRANpkg{PSweight} allows for symmetric propensity score trimming \citep{Crump2009,Yoshida2019} and optimal trimming \citep{Crump2009,Yang2016} to improve the internal validity. We extend the class of balance metrics suggested in \citet{austin2015moving} and \citet{LiThomasLi2018} for binary treatments, and those in \citet{McCaffrey2013} and \citet{li2019propensity} for multiple treatments. In addition, the design module helps describe the weighted target population by providing the information required in the standard ``Table 1'' of a clinical article.  

In addition to the standard inverse probability of treatment weights (IPW), \CRANpkg{PSweight} implements the average treatment effect among the treated (Treated) weights, overlap weights (OW), matching weights (MW) and entropy weights (EW) for both binary \citep{LiGreene13,Mao2018,LiMorganZaslavsky2018,Zhou2020} and multiple treatments \citep{Yoshida2017,li2019propensity}. All weights are members of the family of balancing weights \citep{LiMorganZaslavsky2018}; the last three types of weights target at the subpopulation with improved overlap in the covariates between (or across) treatment groups, similar to the target population in randomized controlled trials \citep{thomas2020overlap,thomas2020using}. Among them, OW achieves optimal balance and estimation efficiency \citep{LiMorganZaslavsky2018,LiThomasLi2018}. We also implement the augmented weighting estimators corresponding to each of the above weighting schemes \citep{Mao2018}. By default, \CRANpkg{PSweight} employs parametric regression models to estimate propensity scores and potential outcomes. Nonetheless, it also allows for propensity scores to be estimated by external machine learning methods including generalized boosted regression models \citep{McCaffrey2013} and super learner \citep{van2007super}, or importing any other propensity or outcome model estimates of interest.

\begin{table}[htbp]
\caption{Comparisons of existing R packages that implement propensity score weighting with discrete treatments. Binary treatments and additive estimands are implemented in all packages, and therefore those two columns are omitted.}
\centering\label{tb:summary}
\scriptsize
\begin{tabular}{p{1cm}cccccccc}
\toprule
 & Multiple & Balance & IPW/ATT & OW/other  & Ratio & Augmented & Nuisance-adj & Optimal \\[-0.6ex]
 & treatments & diagnostics & weights & weights  & estimands & weighting & variance & trimming \\
\midrule
\CRANpkg{PSweight}  & $\checkmark$ & $\checkmark$ & $\checkmark$ & $\checkmark$ & $\checkmark$ & $\checkmark$ & $\checkmark$ & $\checkmark$  \\
\CRANpkg{twang} & $\checkmark$ & $\checkmark$ & $\checkmark$ & $\times$  & $\times$ & $\times$ & $\times$ & $\times$ \\
\CRANpkg{CBPS} & $\checkmark$ & $\checkmark$ & $\checkmark$ & $\times$   & $\times$ & $\checkmark$ & $\checkmark$ & $\times$ \\
\CRANpkg{PSW} & $\times$ & $\checkmark$ & $\checkmark$ & $\checkmark$  & $\checkmark$ & $\checkmark$ & $\checkmark$ & $\times$  \\
\CRANpkg{optweight}  & $\checkmark$ & $\times$ & $\checkmark$ & $\times$  & $\times$ & $\times$ & $\times$ & $\times$ \\
\CRANpkg{ATE} & $\checkmark$ & $\checkmark$ & $\checkmark$ & $\times$  & $\times$ & $\times$ & $\checkmark$ & $\times$ \\
\CRANpkg{WeightIt} & $\checkmark$ & $\times$ & $\checkmark$ & $\checkmark$  & $\times$ & $\times$ & $\times$ & $\times$ \\
\CRANpkg{causalweight}  & $\checkmark$ & $\times$ & $\checkmark$ & $\times$  & $\times$ & $\checkmark$ & $\times$ & $\times$ \\
\CRANpkg{sbw}  & $\times$ & $\checkmark$ & $\checkmark$ & $\times$  & $\times$ & $\times$ & $\times$ & $\times$ \\
\bottomrule
\end{tabular}
\begin{tablenotes}
\item $\checkmark$ indicates that the functionality is currently implemented in the package; $\times$ indicates otherwise. 
\item References:  \CRANpkg{twang} (Version 1.6): \citet{twang}; \CRANpkg{CBPS} (Version 0.21): \citet{CBPS}; \CRANpkg{PSW} (Version 1.1-3): \citet{PSW}; \CRANpkg{optweight} (Version 0.2.5): \citet{optweight};
\CRANpkg{ATE} (Version 0.2.0): \citet{ATE}; \CRANpkg{WeightIt} (Version 0.10.2): \citet{WeightIt}; 
\CRANpkg{causalweight} (Version 0.2.1): \citet{causalweight}; \CRANpkg{sbw} (Version 1.1.1): \citet{sbw}.  
\end{tablenotes}
\end{table}

To our knowledge, \CRANpkg{PSweight} is the first R package to accommodate a variety of balancing weighting schemes with multiple treatments. Existing R packages such as \CRANpkg{twang}~\citep{twang}, \CRANpkg{CBPS}~\citep{CBPS}, \CRANpkg{optweight}~\citep{optweight} have also implemented weighting-based estimation with multiple treatments, but focus on IPW. The \CRANpkg{PSW} R package~\citep{PSW}  implements both OW and MW and allows for nuisance-adjusted variance estimation, but it is only for binary treatments.
To better assist applied researchers to perform propensity score weighting analysis, this article provides a full introduction of the \CRANpkg{PSweight} package. In Section \ref{sec:models}, we explain the methodological foundation of \CRANpkg{PSweight}. Section \ref{sec:overview} outlines the main functions and their arguments. Section \ref{sec:illustrations} illustrates the use of these functions with a data example that studies the causal effect of educational attainment on income. Section \ref{sec:summary} concludes with a short discussion and outlines future development.

\section{Overview of propensity score weighting} \label{sec:models}

Before diving into the implementation details of \CRANpkg{PSweight}, we briefly introduce the basics of the propensity score weighting framework. 

\subsection{Binary treatments}\label{sec:binary}

Assume we have an observational study with $N$ units. Each unit $i$ ($i=1,2,\ldots,N$) has a binary treatment indicator $Z_{i}$ ($Z_{i}=0$ for control and $Z_{i}=1$ for treated), a vector of $p$ covariates $\bX_{i}=(X_{1i},\cdots, X_{pi})$. For each unit $i$, we assume a pair of potential outcomes $\{Y_{i}(1),Y_{i}(0)\}$ mapped to the treatment and control status, of which only the one corresponding to the observed treatment is observed, denoted by $Y_i=Z_{i}Y_{i}(1)+(1-Z_{i})Y_{i}(0)$; the other potential outcome is counterfactual.

Causal effects are contrasts of the potential outcomes of the same units in a \emph{target population}, which usually is the population of a scientific interest \citep{thomas2020using}. \CRANpkg{PSweight} incorporates a general class of weighted average treatment effect (WATE) estimands. Specifically, assume the observed sample is drawn from a probability density $f(\bx)$, and let $g(\bx)$ denote the covariate density of the target population. The ratio $h(\bx)\propto g(\bx)/f(\bx)$ is called the \emph{tilting function}, which adjusts the distribution of the observed sample to represent the target population. Denote the conditional expectation of the potential outcome by $m_z(\bx)=\bE[Y(z)|\bX=\bx]$ for $z=0,1$. Then, we can represent the average treatment effect over the target population by a WATE estimand:
\begin{equation}\label{eq:estimand1}
\tau^h=\bE_g[Y(1)-Y(0)]=\frac{\bE\{h(\bx)(m_1(\bx)-m_0(\bx))\}}{\bE\{h(\bx)\}}.    
\end{equation}
To estimate \eqref{eq:estimand1}, \CRANpkg{PSweight} maintains two standard assumptions: (1) \emph{unconfoundedness}: $\{Y(1),Y(0)\} \perp Z \mid \bX$; (2) \emph{overlap}: $0<P(Z=1|\bX)<1$. 
The propensity score is the probability of a unit being assigned to the treatment group given the covariates \citep{Rosenbaum1983}: $e(\bx)=P(Z=1|\bX=\bx)$. While assumption (1) is generally untestable and critically depends on substantive knowledge, assumption (2) can be checked visually from data by comparing the distribution of propensity scores between treatment and control groups.

For a given tilting function $h(\bx)$ (and correspondingly a WATE estimand $\tau^h$), we can define the \emph{balancing weights} $(w_1,w_0)$ for the treated and control units:  $w_1(\bx) \propto h(\bx)/{e(\bx)}$ and $w_0(\bx) 
\propto h(\bx)/\{1-e(\bx)\}$. These weights balance the covariate distributions between the treated and control groups towards the target population \citep{LiMorganZaslavsky2018}. \CRANpkg{PSweight} implements the following H\'{a}jek estimator for WATE:
\begin{equation}
\label{eq:sampleWATE}
\hat{\tau}^h=\hat{\mu}^h_1-\hat{\mu}^h_0=\frac{\sumi w_1(\bx_i)Z_i Y_i}{\sumi w_1(\bx_i)Z_i} -
              \frac{\sumi w_0(\bx_i)(1-Z_i) Y_i}{\sumi w_0(\bx_i)(1-Z_i)},
\end{equation}
where the weights are calculated based on the propensity scores estimated from the data. Clearly, specification of $h(\bx)$ defines the target population and estimands. \CRANpkg{PSweight} primarily implements the following three types of balancing weights (see Table \ref{tab:weights_binary} for a summary):
\begin{itemize}
    \item \emph{Inverse probability of treatment weights} (IPW), whose target population is  the combined treatment and control group represented by the observed sample, and the target estimand is the average treatment effect among the combined population (ATE).
    \item \emph{Treated weights}, whose target population is the treated group, and target estimand is the average treatment effect for the treated population (ATT). Treated weights can be viewed as a special case of IPW because it inversely weights the control group. 
    \item \emph{Overlap weights} (OW) \citep{LiMorganZaslavsky2018, li2019propensity}, whose target population is the subpopulation with the most overlap in the observed covariates between treatment and group groups . In medicine  this is known as the population in clinical equipoise and is the population eligible to be enrolled in randomized clinical trials.
    The target estimand of OW is the average treatment effect for the overlap population (ATO).
 
\end{itemize}
IPW has been the dominant weighting method in the literature, but has a well-known shortcoming of being sensitive to extreme propensity scores, which induces bias and large variance in estimating treatment effects. OW addresses the conceptual and operational problems of IPW.   
Among all balancing weights, OW leads to the smallest asymptotic (and often finite-sample) variance of the weighting estimator \eqref{eq:sampleWATE}.  \citep{LiMorganZaslavsky2018,LiThomasLi2018}.  
Recent simulations also show that OW provides more stable causal estimates under limited overlap \citep{LiThomasLi2018,Mao2018,Yoshida2017,Yoshida2019}, and is more robust to misspecification of the propensity score model \citep{Zhou2020}. 

\CRANpkg{PSweight} implements two additional types of balancing weights: matching weights (MW) \citep{LiGreene13}, and entropy weights (EW) \citep{Zhou2020}.  Similar to OW, MW and EW focus on target populations with substantial overlap between treatment groups. Though having similar operating characteristics, MW and EW do not possess the same theoretical optimality as OW, and are less used in practice. Therefore, we will not separately describe MW and EW hereafter. 

\begin{table} 
\begin{center}
\caption{Target populations, tilting functions, estimands and the corresponding balancing weights for binary treatments in \CRANpkg{PSweight}. \label{tab:weights_binary}}
{\footnotesize
\begin{threeparttable}
\begin{tabular}{lccc}
\toprule
Target population  &Tilting function $h(\bx)$ & Estimand & Balancing weights $(w_1, w_0)$  \\
\midrule
Combined &1         &ATE   &$\left(\frac{1}{e(\bx)}, \frac{1}{1-e(\bx)}\right)$   \\
Treated  &$e(\bx)$    &ATT   &$\left(1, \frac{e(\bx)}{1-e(\bx)}\right)$   \\
Overlap  &$e(\bx)(1-e(\bx))$ &ATO & $(1-e(\bx), e(\bx))$  \\
Matching &$\xi_1(\bx)$  & ATM & $\left(  \frac{\xi_1(\bx)}{e(\bx)}, \frac{\xi_1(\bx)}{1-e(\bx)} \right)$ \\
Entropy  &$\xi_2(\bx)$ & ATEN & $\left( \frac{ \xi_2(\bx)}{e(\bx)}, \frac{ \xi_2(\bx)}{1-e(\bx)} \right)$ \\
\bottomrule
\end{tabular}
\begin{tablenotes}
      \small
      \item Notes: $\xi_1(\bx) = \min \{e(\bx),1-e(\bx)\}$ and $\xi_2(\bx)=-\{e(\bx)\log(e(\bx))+(1-e(\bx))\log(1-e(\bx))\} $.
\end{tablenotes}
\end{threeparttable}}
\end{center}
\end{table}

In observational studies, propensity scores are generally unknown and need to be estimated. Therefore, propensity score analysis usually involves two steps: (1) estimating the propensity scores, and (2) estimating the causal effects based on the estimated propensity scores. In \CRANpkg{PSweight}, the default model for estimating propensity scores with binary treatments is a logistic regression model. Spline or polynomial models can be easily incorporated by adding \code{bs()}, \code{ns()} or \code{poly()} terms into the model formula. \CRANpkg{PSweight} also allows for importing propensity scores estimated from external routines, such as boosted models or super learner (Section \ref{sec:impillu}).  

Goodness-of-fit of the propensity score model is usually assessed based on the resulting covariate balance. In the context of propensity score weighting, this is  measured based on either the absolute standardized difference (ASD):
\begin{equation}\label{eq:ASD1}
\text{ASD} = \left| {\frac{\sum_{i=1}^N  w_1(\bx_i)Z_iX_{pi} }{\sum_{i=1}^N w_1(\bx_i)Z_i } - \frac{\sum_{i=1}^N  w_0(\bx_i)(1- Z_i)X_{pi} }{\sum_{i=1}^N w_0(\bx_i)(1-Z_i) }}\right|
\Bigg /{\sqrt{\frac{s_{1}^2 + s_{0}^2}{2}}},
\end{equation}
or the target population standardized difference (PSD), $\max\{\text{PSD}_0,\text{PSD}_1\}$, where
\begin{equation}\label{eq:PSD1}
\text{PSD}_z =
\left|{\frac{\sum_{i=1}^N  w_z(\bx_i)\mathds{1}\{Z_i=z\}X_{pi} }{\sum_{i=1}^N w_z(\bx_i)\mathds{1}\{Z_i=z\} } - \frac{\sum_{i=1}^N h(\bx_i)X_{pi} }{\sum_{i=1}^N h(\bx_i)}}\right|\Bigg /{\sqrt{\frac{s_{1}^2 + s_{0}^2}{2}}}.
\end{equation}
In \eqref{eq:ASD1} and \eqref{eq:PSD1}, $s_z^2$ is the variance (either unweighted or weighted, depending on user specification) of the $p$th covariate in group $z$, and $(w_0,w_1)$ are the specified balancing weights. Setting $w_0=w_1=1$ corresponds to the unweighted mean differences. ASD and PSD are often displayed as column in the baseline characteristics table (known as the ``Table 1'') and visualized via a Love plot (also known as a forest plot) \citep{Greifer}. A rule of thumb for determining adequate balance is when ASD of all covariates is controlled within $0.1$ \citep{austin2015moving}.

\subsection{Multiple treatments}
\label{sec:multiple}

\cite{li2019propensity} extend the framework of balancing weights to multiple treatments. Assume that we have $J$ $(J\geq 3)$ treatment groups, and let $Z_i$ stand for the treatment received by unit $i$, $Z_i\in \{1,\ldots,J\}$. We further define $D_{ij}=\mathds{1}\{Z_i=j\}$ as a set of multinomial indicator, satisfying $\sum_{i=1}^J D_{ij}=1$ for all $j$. Denote the potential outcome for unit $i$ under treatment $j$ as $Y_{i}(j)$, of which only the one corresponding to the received treatment, $Y_i=Y_i(Z_i)$, is observed. The generalized propensity score is the probability of receiving a potential treatment $j$ given $\bX$ \citep{Imbens2000}: $e_j(\bx)=P(Z=j|\bX=\bx)$, with the constraint that $\sum_{j=1}^J e_j(\bx)=1$.

To define the target estimand, let $m_j(\bx)=\bE[Y(j)|\bX=\bx]$ be the conditional expectation of the potential outcome in group $j$. For specified tilting function $h(\bx)$ and target density $g(\bx)\propto f(\bx)h(\bx)$, the $j$th average potential outcome among the target population is
\begin{equation} 
\label{eq:meanpo}
\mu_j^h=\bE_g[Y(j)]=\frac{\bE\{h(\bx)m_j(\bx)\}}{\bE\{h(\bx)\}}.
\end{equation}
Causal estimands can then be constructed in a general manner as contrasts based on $\mu_j^h$. For example, the most commonly seen estimands in multiple treatments are the pairwise average treatment effects between groups $j$ and $j'$: $\tau_{j,j'}^h=\mu_j^h-\mu_{j'}^h$. This definition can be generalized to arbitrary linear contrasts. Denote $\pmb{a}=(a_{i},\cdots, a_{J})$ as a contrast vector of length $J$. A general class of additive estimands is 
\begin{equation} \label{eq:meantau}
\tau^h(\pmb{a})=\sum\limits_{j=1}^{J}a_j\mu_j^h.    
\end{equation}
Specific choices for $\bm{a}$ with nominal and ordinal treatments can be found in \citet{li2019propensity}. Similar as before, propensity score weighting analysis with multiple treatments rests on two assumptions: (1) \emph{weak unconfoundedness}: $Y(j)\perp \mathds{1}\{Z=j\} |\bX,$ for all $j$, and (2) \emph{Overlap}: the generalized propensity score is bounded away from 0 and 1: $0<e_j(\bx)<1$, for all $j$.

With multiple treatments, the tilting function $h(\bx)$ specifies the target population, estimand, and balancing weights. For a given $h(\bx)$, the balancing weights for the $j$th treatment group  $w_j(\bx)\propto {h(\bx)}/{e_j(\bx)}$. Then the H\'{a}jek estimator for $\mu_j^h$ is 
\be
\label{eq:simplemean}
\hat{\mu}^h_j=\frac{\sum_{i=1}^N w_j(\bx_i)D_{ij}Y_i  }{\sum_{i=1}^N w_j(\bx_i)D_{ij}}.
\ee
Contrasts based on $\hat{\mu}^h_j$ can be obtained for any $\ba$ to estimate the additive causal estimand $\tau^h(\ba)$. Of note, we only consider types of estimands that are transitive, and therefore the ATT estimands introduced in \citet{Lechner2001} is not implemented. In parallel to binary treatments \CRANpkg{PSweight} implements five types of balancing weights with multiple treatments: IPW, treated weights, OW, MW, and EW, and the corresponding target estimand of each weighting scheme is its pairwise (between each pair of treatments) counterpart in binary treatments.  
 
Among all the weights, OW minimizes the total asymptotic variances of all pairwise comparisons, and has been shown to have the best finite-sample efficiency in estimating pairwise WATEs \citep{li2019propensity}.  Table \ref{tab:weight_multi} summarizes the target population, tilting function and balancing weight for multiple treatments that are available in \CRANpkg{PSweight}.

\begin{table}[htbp]
\centering
\caption{Target populations, tilting functions, and the corresponding balancing weights for multiple treatments in \CRANpkg{PSweight}.}\label{tab:weight_multi}
{\footnotesize
\begin{tabular}{lcc}
\toprule
 Target population & Tilting function $h(\bx)$ & Balancing weights $\left\{w_j(\bx),~j=1,\ldots,J\right\}$\\\midrule
Combined & 1 & $\left\{1/e_j(\bx)\right\}$\\
Treated ($j'$th group) & $e_{j'}(\bx)$ & $\left\{e_{j'}(\bx)/e_j(\bx)\right\}$ \\
Overlap & $\{\sumk  1/e_k(\bx)\}^{-1}$ & $\left\{\{\sumk  1/e_k(\bx)\}^{-1}/e_j(\bx)\right\}$\\
Matching &$\min_k \{e_k(\bx)\}$  & $\left\{ \min_k \{e_k(\bx)\}/e_j(\bx) \right\}$ \\
Entropy  &$-\sum_{k=1}^J e_k(\bx)\log\{e_k(\bx)\}$ & $\left\{-\sum_{k=1}^J e_k(\bx)\log\{e_k(\bx)\}/ e_j(\bx) \right\}$ \\
\bottomrule
\end{tabular}}
\end{table}

To estimate the generalized propensity scores for multiple treatments, the default model in \CRANpkg{PSweight} is a multinomial logistic model. \CRANpkg{PSweight} also allows for externally estimated generalized propensity scores. Goodness-of-fit of the generalized propensity score model is assessed by the resulting covariate balance, which is measured by the pairwise versions of the ASD and PSD. The detailed formula of these metrics can be found in \cite{li2019propensity}. A common threshold for balance is that the maximum pairwise ASD or maximum PSD is below 0.1.  
\subsection{Propensity score trimming}

Propensity score trimming excludes units with estimated (generalized) propensity scores close to zero (or one). It is a popular approach to address the extreme weights problem of IPW. \CRANpkg{PSweight} implements the symmetric trimming rules in \citet{Crump2009} and \citet{Yoshida2019}.  Operationally, we allow users to specify a single cutoff $\delta$ on the estimated generalized propensity scores, and only includes units for analysis if $\min_j\{e_{j}(\bx)\}\in[\delta,1]$. With binary treatments, the symmetric trimming rule reduces to $e(\bx)\in[\delta,1-\delta]$. The natural restriction $\delta<1/J$ must be satisfied due to the constraint  $\sum_{j=1}^J e_j(\bx)=1$.   
To avoid specifying an arbitrary trimming threshold $\delta$, \CRANpkg{PSweight} also implements the optimal trimming rules of \citet{Crump2009} and \citet{Yang2016}, which minimizes the (total) asymptotic variance(s) for estimating the (pairwise) ATE among the class of all trimming rules. OW can be viewed as a continuous version of trimming because it smoothly down-weigh the units with propensity scores close to 0 or 1, and thus avoids specifying a threshold.

\subsection{Augmented weighting estimators}
\label{sec:augest}

\CRANpkg{PSweight} also implements augmented weighting estimators, which augment a weighting estimator by an outcome regression and improves the efficiency.  With IPW, the augmented weighting estimator is known as the doubly-robust estimator \citep{Lunceford2004,bang2005doubly,funk2011doubly}. With binary treatments, the augmented estimator with general balancing weights are discussed \citet{Hirano2003} and \citet{Mao2018}. Below, we briefly outline the form of this estimator with multiple treatments. Recall the conditional mean of $Y_i(j)$ given $\bX_i$ and treatment $Z_i=j$ as $m_{j}(\bx_i)=\bE[Y_i(j)|\bX_i=\bx_i]=\bE[Y_i|\bX_i=\bx_i,Z_{i}=j]$.  
This conditional mean can be estimated by generalized linear models, kernel estimators, or machine learning models. \CRANpkg{PSweight} by default employs the generalized linear models, but also allows estimated values from other routines. When $m_{j}(\bx_i)$ is estimated by generalized linear models, \CRANpkg{PSweight} currently accommodates three types of outcomes: continuous, binary and count outcomes (with or without an offset), using the canoncal link function.  

With a pre-specified tilting function, the augmented weighting estimator for group $j$ is 
\be\label{eq:augest}
\hat{\mu}^{h,\aug}_j=\frac{\sum_{i=1}^N w_j(\bx_i)D_{ij}\{Y_i-m_{j}(\bx_i)\}  }{\sum_{i=1}^N w_j(\bx_i)D_{ij}}+\frac{\sum_{i=1}^N h(\bx_i)m_{j}(\bx_i)  }{\sum_{i=1}^N h(\bx_i)}.
\ee
The first term of (\ref{eq:augest}) is the H\'{a}jek estimator of the regression residuals, and the second term is the standardized average potential outcome (a $g$-formula estimator). With IPW, (\ref{eq:augest}) is consistent to $\bE[Y(j)]$ when either the propensity score model or the outcome model is correctly specified, but not necessarily both. For other balancing weights, (\ref{eq:augest}) is consistent to the WATE when the propensity model is correctly specified, regardless of outcome model specification. When both models are correctly specified, (\ref{eq:augest}) achieves the lower bound of the variance for regular and asymptotic linear estimators \citep{Robins1994,Hirano2003,Mao2018}.

\subsection{Ratio causal estimands}\label{sec:ratioest}

With binary and count outcomes, ratio causal estimands are often of interest. Using notation from the multiple treatments as an example, once we use weighting to obtain estimates for the set of average potential outcomes $\{\mu_j^h,j=1,\ldots,J\}$, we can directly estimate the causal relative risk (RR) and causal odds ratio (OR), defined as
\begin{equation}\label{eq:RROR}
\tau^{h,\RR}_{j,j^{\prime}}=\frac{\mu_{j}^h}{\mu_{j^{\prime}}^h},~~~~~~\tau^{h,\OR}_{j,j^{\prime}}=\frac{\mu_{j}^h/(1-\mu_{j}^h)}{\mu_{j^{\prime}}^h/(1-\mu_{j^{\prime}}^h)}.    
\end{equation}
Here the additive estimand $\tau^{h,\RD}_{j,j^{\prime}}=\mu_{j}^h-\mu_{j^{\prime}}^h$ is the causal risk difference (RD). \CRANpkg{PSweight} supports a class of ratio estimands for any given contrasts $\ba$. Specifically, we define the log-RR type parameters by
\begin{equation} \label{eq:meanRR}
\lambda^{h,\RR}(\pmb{a})=\sum\limits_{j=1}^{J}a_j\log\left(\mu_j^h\right),
\end{equation}
and the log-OR type parameters by
\begin{equation} \label{eq:meanOR}
\lambda^{h,\OR}(\pmb{a})=\sum\limits_{j=1}^{J}a_j\left\{\log\left(\mu_j^h\right)-\log\left(1-\mu_j^h\right)\right\}.
\end{equation}
With nominal treatments, the contrast vector $\ba$ can be specified to encode pairwise comparisons in the log scale (as in \eqref{eq:meanRR}) or in the log odds scale (as in \eqref{eq:meanOR}), in which case $\exp\{\lambda^{h,\RR}(\pmb{a})\}$ and $\exp\{\lambda^{h,\OR}(\pmb{a})\}$ become the causal RR and causal OR in \eqref{eq:RROR}.  
User-specified contrasts $\ba$ can provide a variety of nonlinear estimands. For example, when $J=3$, with $\ba=(1,-2,1)^T$ one can use \CRANpkg{PSweight} to assess the equality of two consecutive causal RR: $H_0: \mu_3^h/\mu_2^h=\mu_2^h/\mu_1^h$.

\subsection{Variance and interval estimation}\label{sec:variance}

\CRANpkg{PSweight} by default implements the empirical sandwich variance for propensity score weighting estimators \citep{Lunceford2004,LiThomasLi2018,Mao2018}  based on the M-estimation theory \citep{Stefanski2002}. The variance adjusted for the uncertainty in estimating the propensity score and outcome models, and are sometime referred to as the nuisance-adjusted sandwich variance. Below we illustrate the main steps with multiple treatments and general balancing weights. Write $\btheta=\left(\nu_1,\ldots,\nu_J,\eta_1,\ldots,\eta_J,\bbeta^T,\balpha^T\right)^T$ as the collection of parameters to be estimated. Then $\left\{\hat{\mu}^{h,\aug}_j=\hat{\nu}_j+\hat{\eta}_j:j=1,\ldots,J\right\}$ jointly solve 
\begin{align*}
\sum_{i=1}^{N}\Psi_{i}(\btheta)=\sum_{i=1}^{N}
\left(
\begin{array}{c}
 w_1(\bx_i)D_{i1}\{Y_{i}-m_{1}(\bx_{i};\balpha)-\nu_1\}\\
\vdots\\ 
w_J(\bx_i)D_{iJ}\{Y_{i}-m_{J}(\bx_{i};\balpha)-\nu_J\}\\
h(\bx_i)\{m_{1}(\bx_{i};\balpha)-\eta_1\}\\
\vdots\\ 
h(\bx_i)\{m_{J}(\bx_{i};\balpha)-\eta_J\}\\
S_{\bbeta}(Z_i,\bx_{i};\bbeta)\\ 
S_{\balpha}(Y_i,Z_i,\bx_{i};\balpha)
\end{array}
\right)=\bm{0},
\end{align*}
where $S_{\bbeta}(Z_i,\bx_{i};\bbeta)$ and $S_{\balpha}(Y_i,Z_i,\bx_{i};\balpha)$ are the score functions of the propensity score model and the outcome model. The empirical sandwich variance estimator is  
\bee
\widehat{\bV}(\hat{\btheta})=\left\{\sum_{i=1}^{N}
\frac{\partial}{\partial\btheta^T}\Psi_{i}(\hat{\btheta})\right\}^{-1} \left\{\sum_{i=1}^{N}\Psi_{i}(\hat{\btheta})
\Psi_{i}^{T}(\hat{\btheta})\right\}
\left\{\sum_{i=1}^{N}
\frac{\partial}{\partial\btheta}\Psi_{i}^T(\hat{\btheta})\right\}^{-1}.
\eee
Because $\hat{\mu}^{h,\aug}_j=\hat{\nu}_j+\hat{\eta}_j$, the variance of arbitrary linear contrasts based on the average potential outcomes can be easily computed by applying the Delta method to the joint variance $\widehat{\bV}(\hat{\btheta})$. For the H\'ajek weighting estimators, variance is estimated by removing $S_{\balpha}(Y_i,Z_i,\bx_{i};\balpha)$ as well as the components involving $m_j(\bx_i;\balpha)$ in $\Psi_{i}(\btheta)$. Finally, when propensity scores and potential outcomes are not estimated through the generalized linear model or are supplied externally,  or MW are used (since the tilting function is not everywhere differentiable), \CRANpkg{PSweight} ignores  the uncertainty in estimating $\bbeta$ and $\balpha$ and removes $S_{\bbeta}(Z_i,\bx_{i};\bbeta)$ and $S_{\balpha}(Y_i,Z_i,\bx_{i};\balpha)$ in $\Psi_{i}(\btheta)$ in the calculation of the empirical sandwich variance. Based on the estimated variance, \CRANpkg{PSweight} computes the associated symmetric confidence intervals and p-values via the normal approximation.   

For ratio causal estimands, \CRANpkg{PSweight} applies the logarithm transformation to improve the accuracy of the normal approximation \citep{agresti2003categorical}. For estimating the variance of causal RR, we first obtain the joint variance of $\left(\log\left(\hat{\mu}^{h,\aug}_1\right),\ldots \log\left(\hat{\mu}^{h,\aug}_J\right)\right)^T$ using the Delta method, and then estimate the variance of $\lambda^{h,\RR}(\pmb{a})$. Once the symmetric confidence intervals are obtained for $\lambda^{h,\RR}(\pmb{a})$ using the normal approximation, we can exponentiate the upper and lower confidence limits to derive the asymmetric confidence intervals for the causal RR. Confidence intervals for the causal OR are computed similarly.

\CRANpkg{PSweight} also allows using bootstrap to estimate variances, which can be much more computationally intensive than the closed-form sandwich estimator but sometimes give better finite-sample performance in small samples. By default, \CRANpkg{PSweight} resamples $R=50$ bootstrap replicates with replacement. For each replicate, the weighting estimator (\ref{eq:simplemean}) or the augmented weighting estimtor (\ref{eq:augest}) is implemented, providing $R$ estimates of the $J$ average potential outcomes (an $R\times J$ matrix). Then for any contrast vector $\pmb{a}=(a_{1},\cdots, a_{J})^T$, \CRANpkg{PSweight} obtains $R$ bootstrap estimates:
\begin{equation*}
  \hat{\mathbb{T}}^h(\pmb{a})_{bootstrap}=  \left\{\hat{\tau}^h(\pmb{a})_1=\sum_{j=1}^{J} a_{j} \hat{\mu}^h_{j,1},~\ldots~, \hat{\tau}^h(\pmb{a})_R=\sum_{j=1}^{J} a_{j} \hat{\mu}^h_{j,R} \right\}.
\end{equation*}
The sample variance of $\hat{\mathbb{T}}^h(\pmb{a})_{bootstrap}$ is reported by \CRANpkg{PSweight} as the bootstrap variance; the lower and upper $2.5\%$ quantiles of $\hat{\mathbb{T}}^h(\pmb{a})_{bootstrap}$ form the $95\%$ bootstrap interval estimate

\section{Overview of package} \label{sec:overview}
The \CRANpkg{PSweight} package includes two modules tailored for design and analysis of observational studies. The design module provides diagnostics to assess the adequacy of the propensity score model and the weighted target population, prior to the use of outcome data. The analysis module provides functions to estimate the causal estimands discussed in Section \ref{sec:models}. We briefly describe the two modules below.

\subsection{Design module}

\CRANpkg{PSweight} offers the \code{SumStat()} function to visualize the distribution of the estimated propensity scores, to assess the balance of covariates under different weighting schemes, and to characterize the weighted target population. It uses the following code snippet:

\begin{Scode}
SumStat(ps.formula, ps.estimate = NULL, trtgrp = NULL, Z = NULL, covM = NULL, 
    zname = NULL,  xname = NULL, data = NULL,weight = "overlap", delta = 0,
    method = "glm", ps.control = list())
\end{Scode}

By default, the (generalized) propensity scores are estimated by the (multinomial) logistic regression, through the argument \code{ps.formula}. Alternatively, \code{gbm()} functions in the \CRANpkg{gbm} package \citep{gbmpkg} or the \code{SuperLearner()} function in the \CRANpkg{SuperLearner} package \citep{slpkg} can also be called by using \code{method = "gbm"} or \code{method = "SuperLearner"}. Additional parameters of those functions can be supplied through the \code{ps.control} argument. The argument \code{ps.estimate} supports estimated propensity scores from external routines. \code{SumStat()} produces a \code{SumStat} object, with estimated propensity scores, unweighted and weighted covariate means for each treatment group, balance diagnostics, and effective sample sizes (defined in \citet{li2019propensity}). We then provide a \code{summary.SumStat()} function, which takes the \code{SumStat} object and summarizes weighted covariate means by treatment groups and the between-group differences in either ASD or PSD. The default options in \code{weighted.var = TRUE} and \code{metric = "ASD"} yield ASD based on weighted standard deviations in \cite{austin2015moving}. The weighted covariate means can be used to build a baseline characteristics ``Table 1'' to illustrate the target population where trimming or balancing weights are applied.

\begin{Scode}
summary(object, weighted.var = TRUE, metric = "ASD")
\end{Scode}

\begin{table}[htbp]
\centering
\caption{\label{tab:design} Functions in the design module of \CRANpkg{PSweight}.}
{\footnotesize
\begin{tabular}{p{4cm}p{8cm}}
\toprule
 Function & Description \\\midrule
 \code{SumStat()} &  Generate a \code{SumStat} object with information of propensity scores and weighted covariate balance \\
\code{summary.SumStat()} & Summarize the \code{SumStat} object and return weighted covariate means by treatment groups and weighted or unweighted between-group differences in ASD or PSD \\
\code{plot.SumStat()} & Plot the distribution of propensity scores or weighted covariate balance metrics from the \code{SumStat} object\\
\code{PStrim()} & Trim the data set based on estimated propensity scores \\
\bottomrule
\end{tabular}}
\end{table}

Diagnostics of propensity score models can be visualized with the \code{plot.SumStat()} function. It takes the \code{SumStat} object and produces a balance plot (\code{type = "balance"}) based on the ASD and PSD. A vertical dashed line can be set by the \code{threshold} argument, with a default value equal to $0.1$. The \code{plot.SumStat()} function can also supply density plot (\code{type = "density"}), or histogram (\code{type = "hist"}) of the estimated propensity scores. The histogram, however, is only available for the binary treatment case. The plot function is implemented as follows:  

\begin{Scode}
plot(x, type = "balance", weighted.var = TRUE, threshold = 0.1, metric = "ASD")
\end{Scode}

In the design stage, propensity score trimming can be carried out with the \code{PStrim()} function. The trimming threshold \code{delta} is set to 0 by default. \code{PStrim()} also enables optimal trimming rules (\code{optimal = TRUE}) that give the most statistically efficient (pairwise) subpopulation ATE, among all possible trimming rules. A trimmed data set along with a summary of trimmed cases will be returned by \code{PStrim()}. This function is given below:

\begin{Scode}
PStrim(data, ps.formula = NULL, zname = NULL, ps.estimate = NULL, delta = 0, 
    optimal = FALSE, method = "glm", ps.control = list())
\end{Scode}

Alternatively, trimming is also anchored in the \code{SumStat()} function with the \code{delta} argument. All functions in the design module are summarized in Table \ref{tab:design}. 

\subsection{Analysis module}

The analysis module of \CRANpkg{PSweight} includes two functions: \code{PSweight()} and \code{summary.PSweight()}. The \code{PSweight()} function estimates the average potential outcomes in the target population, $\{\mu_{j}^{h},j=1,\ldots,J\}$, and the associated variance-covariance matrix. By default, the empirical sandwich variance is implemented, but bootstrap variance can be obtained with the argument \code{bootstrap = TRUE)}. The \code{weight} argument can take \code{"IPW"}, \code{"treated"}, \code{"overlap"}, \code{"matching"} or \code{"entropy"}, corresponding to the weights introduced in Section \ref{sec:models}. More detailed descriptions of each input argument in the \code{PSweight()} function can be found in Table \ref{tab:est}. A typical \code{PSweight()} code snippet looks like

\begin{Scode}
PSweight(ps.formula, ps.estimate, trtgrp, zname, yname, data, weight = "overlap", 
    delta = 0, augmentation = FALSE, bootstrap = FALSE, R = 50, out.formula = NULL,
    out.estimate = NULL, family = "gaussian", ps.method = "glm", ps.control = list(),
    out.method = "glm", out.control = list())
\end{Scode}
 
\begin{table}[!htbp]
\centering
\caption{Arguments for function \code{PSweight()} in the analysis module of \CRANpkg{PSweight}. \label{tab:est}}
{\footnotesize
\begin{tabular}{p{2cm}p{9.5cm}p{2cm}}
\toprule
 Argument & Description & Default \\\midrule
\code{ps.formula} &  A symbolic description of the propensity score model. & --  \\
\code{ps.estimate} & An optional matrix or data frame with externally estimated (generalized) propensity scores for each observation; can also be a vector with binary treatments. & \code{NULL} \\
\code{trtgrp} & An optional character defining the \emph{treated} population for estimating (pairwise) ATT. It can also be used to specify the treatment level when only a vector of values are supplied for \code{ps.estimate} in the binary treatment setting. & Last value in alphabetic order \\
\code{zname} & An optional character specifying the name of the treatment variable when \code{ps.formula} is not provided. & \code{NULL} \\
\code{yname}& A character specifying name of the outcome variable in \code{data}. \\
\code{weight} & A character specifying the type of weights to be used. & \code{"overlap"}\\
\code{delta} & Trimming threshold for (generalized) propensity scores. & 0\\
\code{augmentation} & Logical value of whether augmented weighting estimators should be used. & \code{FALSE} \\
\code{bootstrap} & Logical value of whether bootstrap is used to estimate the standard error  & \code{FALSE}\\
\code{R} & Number of bootstrap replicates if \code{bootstrap = TRUE} & 50\\
\code{out.formula} & A symbolic description of the outcome model to be estimated  when \code{augmentation = TRUE} &\\
\code{out.estimate} & An optional matrix or data frame containing externally estimated potential outcomes for each observation under each treatment level.& \code{NULL} \\
\code{family} & A description of the error distribution and canonical link function to be used in the outcome model if \code{out.formula} is provided & \code{"gaussian"}\\
\code{ps.method}& a character to specify the method for propensity model.
 &\code{"glm"}\\
\code{ps.control}&
A list to specify additional options when \code{method} is set to \code{"gbm"} or \code{"SuperLearner"}.&\code{list()}\\
\code{out.method}& A character to specify the method for outcome model.& \code{"glm"}\\
\code{out.control}&A list to specify additional options when \code{methodout} is set to \code{"gbm"} or \code{"SuperLearner"}. &\code{list()}\\
\bottomrule
\end{tabular}}
\end{table}

Similar to the design module, the \code{summary.PSweight()} function synthesizes information from the \code{PSweight} object for statistical inference. A typical code snippet looks like

\begin{Scode}
summary(object, contrast, type = "DIF", CI = TRUE)
\end{Scode}

In the \code{summary.PSweight()} function, the argument \code{type} corresponds to the three types estimands: \code{type = "DIF"} 
is the default argument that specifies the additive causal contrasts; \code{type = "RR"} specifies the contrast on the log scale as in equation \eqref{eq:meanRR}; \code{type = "OR"} specifies the contrast on the log odds scale as in equation \eqref{eq:meanOR}. Confidence intervals and p-values are obtained using normal approximation and reported by the \code{summary.PSweight()} function. The argument \code{contrast} represents a contrast vector $\pmb{a}$ or matrix with multiple contrast row vectors. If \code{contrast} is not specified, \code{summary.PSweight()} provides all pairwise comparisons of the average potential outcomes. By default, confidence interval is printed (\code{CI = TRUE}); alternatively, one can print the test statistics and p-values by \code{CI = FALSE}.

\section{Case study with the NCDS data} \label{sec:illustrations}

We demonstrate \CRANpkg{PSweight} in a case study that estimates the causal effect of educational attainment on hourly wage, based on the National Child Development Survey (NCDS) data.  
The National Child Development Survey (NCDS) is a longitudinal study on children born in the United Kingdom (UK) in $1958$ \footnote{https://cls.ucl.ac.uk/cls-studies/1958-national-child-development-study/}. NCDS collected information such as educational attainment, familial backgrounds, and socioeconomic and health well being on $17,415$ individuals. We followed \citet{battistin2011misclassified} to pre-process the data and obtain a subset of 3,642 males employed in 1991 with complete educational attainment and wage information for analysis. For illustration, we use the Multiple Imputation by Chained Equations in \citep{buuren2010mice} to impute missing covariates and obtain a single imputed data set for all subsequent analysis.\footnote{Ten out of twelve pre-treatment covariates we considered have missingness. The smallest missingness proportion is $4.9\%$ and the largest missingness proportion is $17.2\%$. We considered one imputed complete data set for illustrative purposes, but note that a more rigorous analysis could proceed by combining analyses from multiple imputed data sets via the Rubin's rule.} The outcome variable \code{wage} is log of the gross hourly wage in Pound. The treatment variable is educational attainment.  
For the multiple treatment case, 
To start with, we created \code{Dmult} as a treatment variable with three levels: \code{">=A/eq"}, \code{"O/eq"} and \code{"None"}, representing advanced qualification ($1,806$ individuals), intermediate qualification ($941$ individuals) and no qualification ($895$ individuals). We consider twelve pre-treatment covariates or potential confounders. The variable \code{white} indicates whether an individual identified himself as white race; \code{scht} indicates the school type they attended at age $16$; \code{qmab} and \code{qmab2} are math test scores at age $7$ and $11$;  \code{qvab} and \code{qvab2} are two reading test scores at age $7$ and $11$; \code{sib$\_$u} stands for the number of siblings; \code{agepa} and \code{agema} are the ages of parents in year 1,974; in the same year, the employment status of mother \code{maemp} was also collected; \code{paed$\_$u} and \code{maed$\_$u} are the years of education for parents. For simplicity, we will focus on IPW and the three types of weights that improve covariate overlap: OW, MW and EW \citep{li2019propensity}.

\subsubsection{Estimating generalized propensity scores and balance assessment}

We use \code{Dmult}, the three-level variable, as the treatment of interest. About one half of the population attained advanced academic qualification, the there are approximately equal number of individuals with intermediate academic qualification or no academic qualification. To illustrate the estimation and inference for ratio estimands, we also introduce a binary outcome of wage, \code{wagebin}. The dichotomized wage was obtained with the cutoff of the average hourly wage of actively employed British male aged $30$-$39$ in $1991$\footnote{https://www.ons.gov.uk/employmentandlabourmarket/peopleinwork/earningsandworkinghours/}. The averaged hourly wage is $8.23$, and we take $\log(8.23)\approx 2.10$ as the cutoff. Among the study participants, we observe $1610$ and $2032$ individuals above and below the average, and we are interested in estimating the pairwise (weighted) average treatment effect of the academic qualification for obtaining above-average hourly wage.

We specify a multinominal regression model, \code{ps.mult}, to estimate the generalized propensity scores.

\begin{Scode}
ps.mult <- Dmult ~ white + maemp + as.factor(scht) + as.factor(qmab)
    as.factor(qmab2) + as.factor(qvab) + as.factor(qvab2) + paed_u + maed_u + 
    agepa + agema + sib_u + paed_u * agepa + maed_u * agema
\end{Scode}
Then we obtain the propensity score estimates and assess weighted covariate balance with the \code{SumStat()} function. 

\begin{Scode}
bal.mult <- SumStat(ps.formula = ps.mult, + 
    weight = c("IPW", "overlap", "matching", "entropy"), data = NCDS)
plot(bal.mult, type = "density")
\end{Scode}

\begin{figure}[!ht]
\centering
  \subfigure{\label{fig:mult1}
  \includegraphics[width=.35\linewidth]{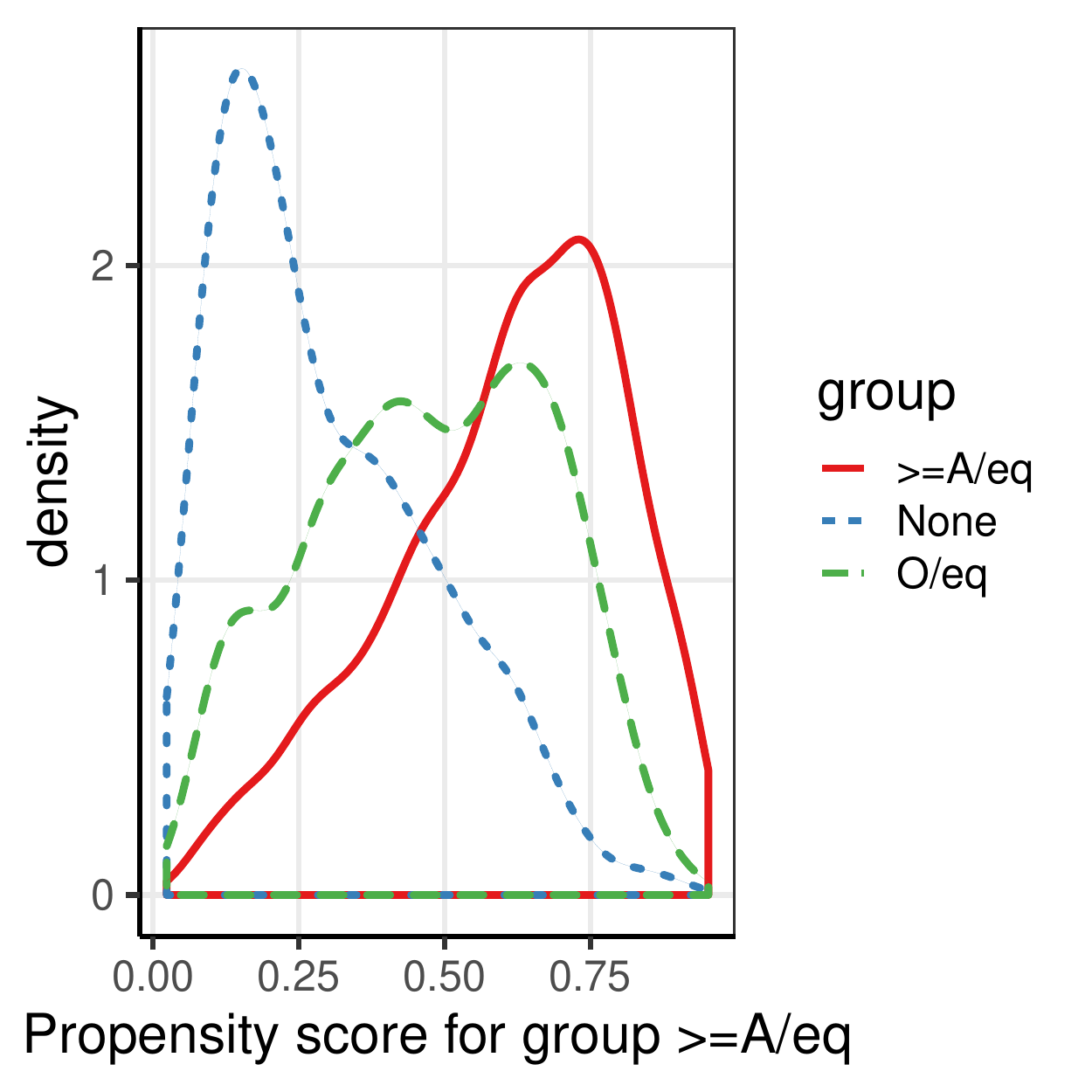} }
    \subfigure{\label{fig:mult2}
    \includegraphics[width=.35\linewidth]{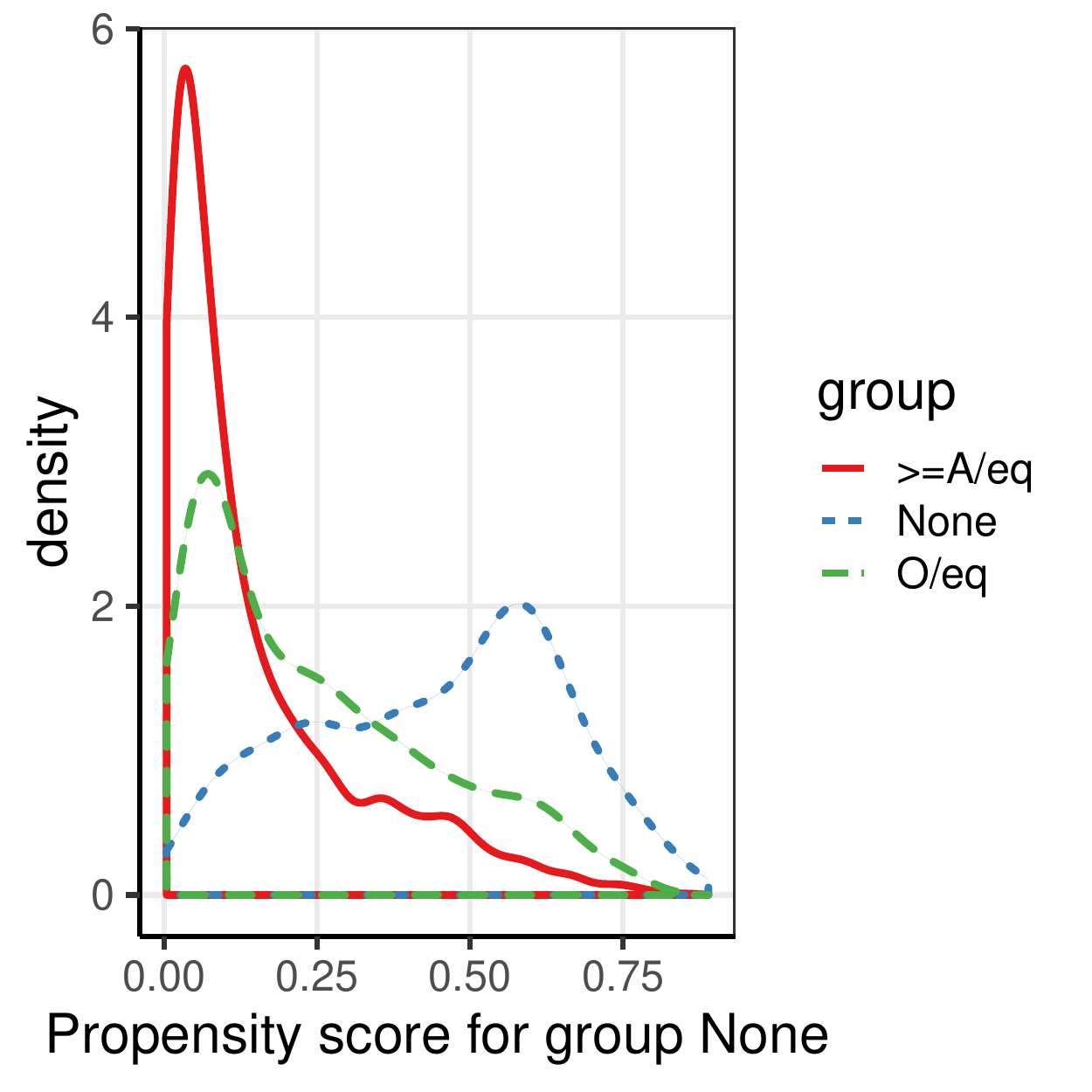} }
    \hspace{-5mm}
        \subfigure{\label{fig:mult3}
    \includegraphics[width=.35\linewidth]{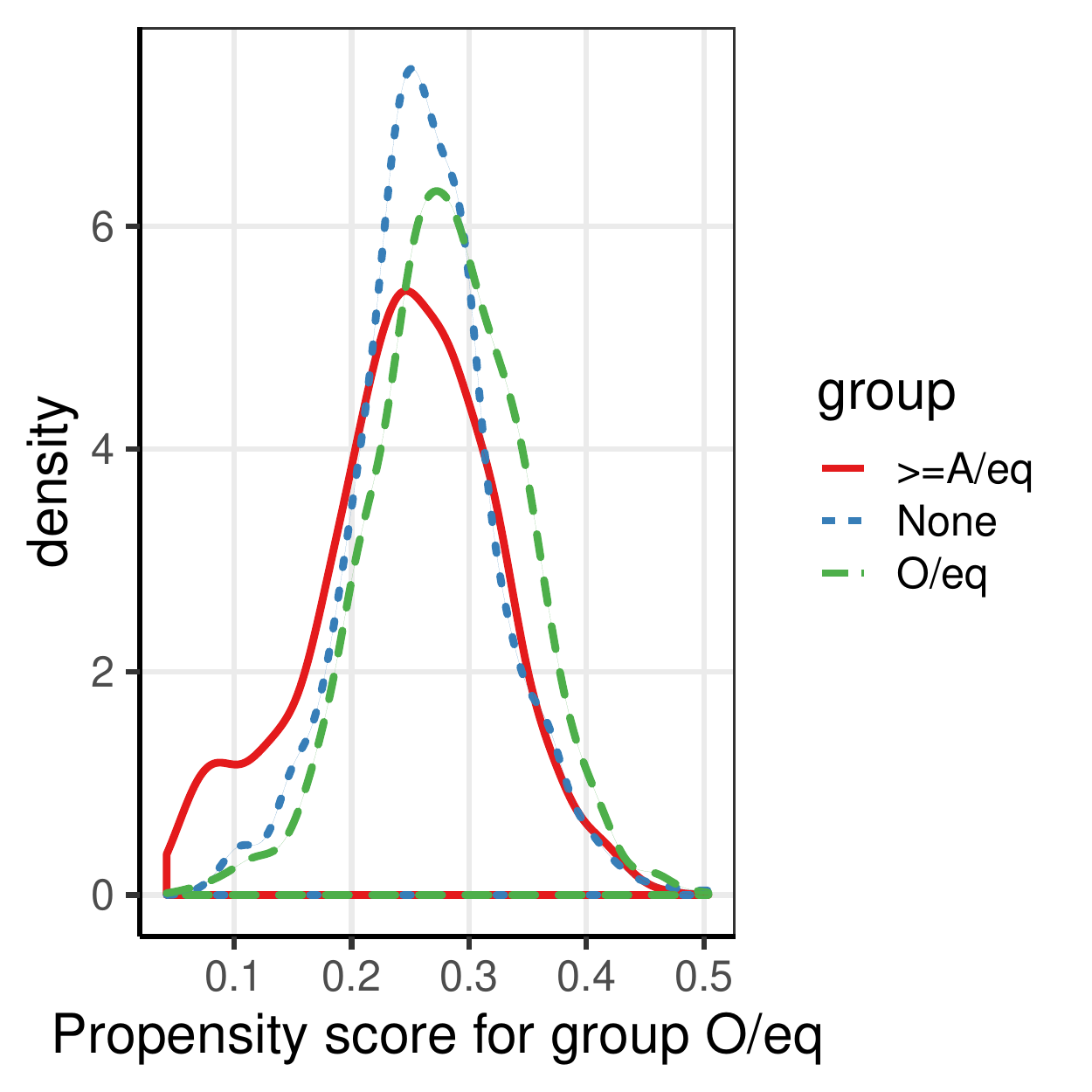}}
\caption{\label{fig:propdmult}
Density plots of estimated generalized propensity scores with respect to the three-level treatment variable \code{Dmult} generated by \code{plot.SumStat()} function in the \CRANpkg{PSweight} package.}
\end{figure}

The distributions of generalized propensity scores are given in Figure \ref{fig:propdmult} (in alphabetic order of the names of treatment groups). For the generalized propensity score to receive the advanced qualification (\code{">=A/eq"}) or no qualification (\code{"None"}), there is a mild lack of overlap due to separation of the group-specific distribution. Since \code{bal.mult} includes four weighting schemes, we plot the maximum pairwise ASD and assess the (weighted) covariate balance in a single Love plot. 

\begin{Scode}
plot(bal.mult, metric = "ASD")
\end{Scode}

The covariates are imbalanced across the three groups prior to any weighting. Although IPW can generally improve covariate balance, the maximum pairwise ASD still ocassionally exceeds the threshold $0.1$ due to lack of overlap. In contrast, OW, MW and EW all emphasize the subpopulation with improved overlap and provide better balance across all covariates. 

\begin{figure}[htbp]
\centering
\includegraphics[width=0.8\textwidth]{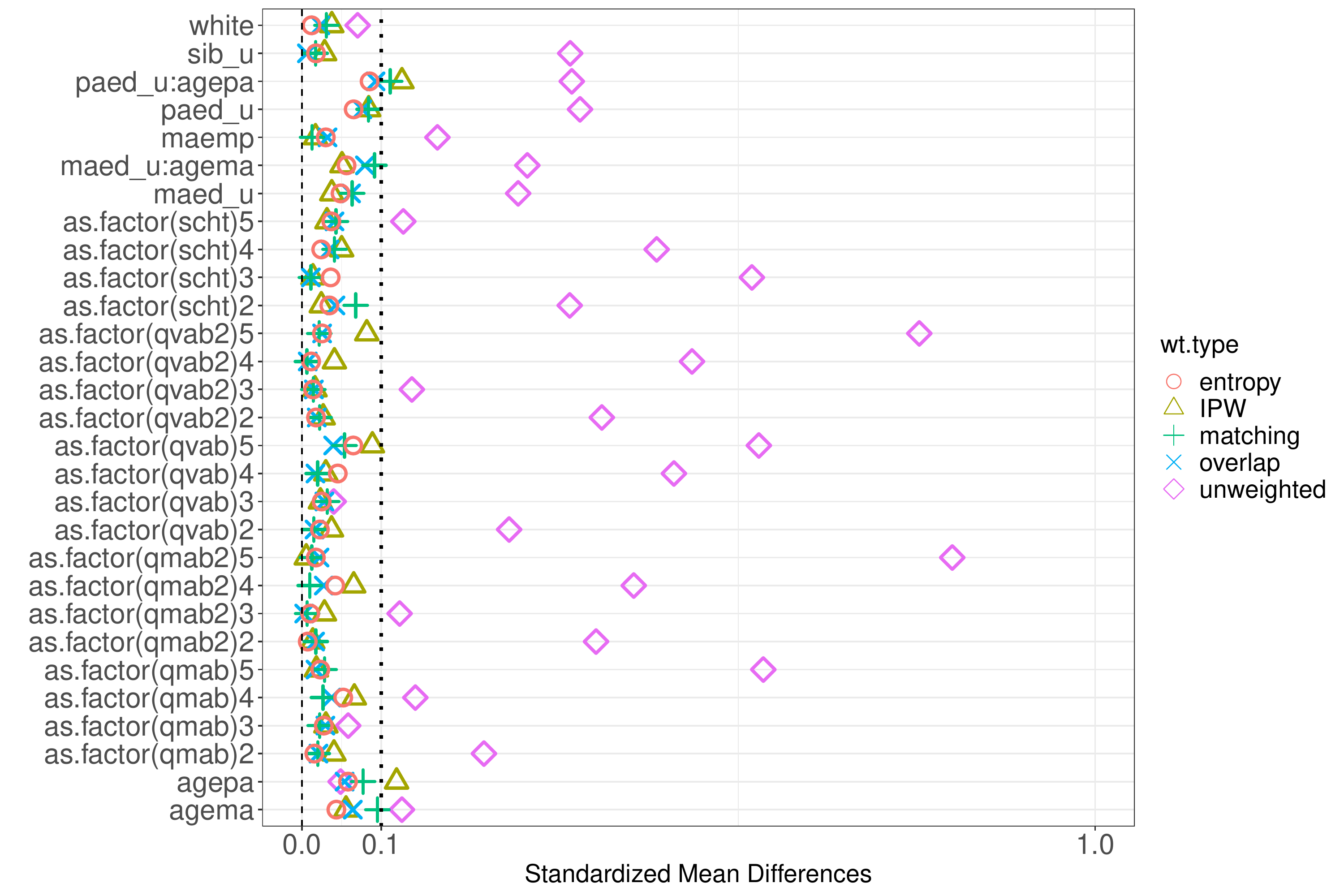}
\caption{Love plot with the three-level treatment variable \code{Dmult} using the maximum pairwise ASD metric, generated by \code{plot.SumStat()} function in the \CRANpkg{PSweight} package.}\label{fig:multasd}
\end{figure}

\subsubsection{Generalized propensity score trimming}

\begin{figure}[h]
\centering
\includegraphics[width=0.8\textwidth]{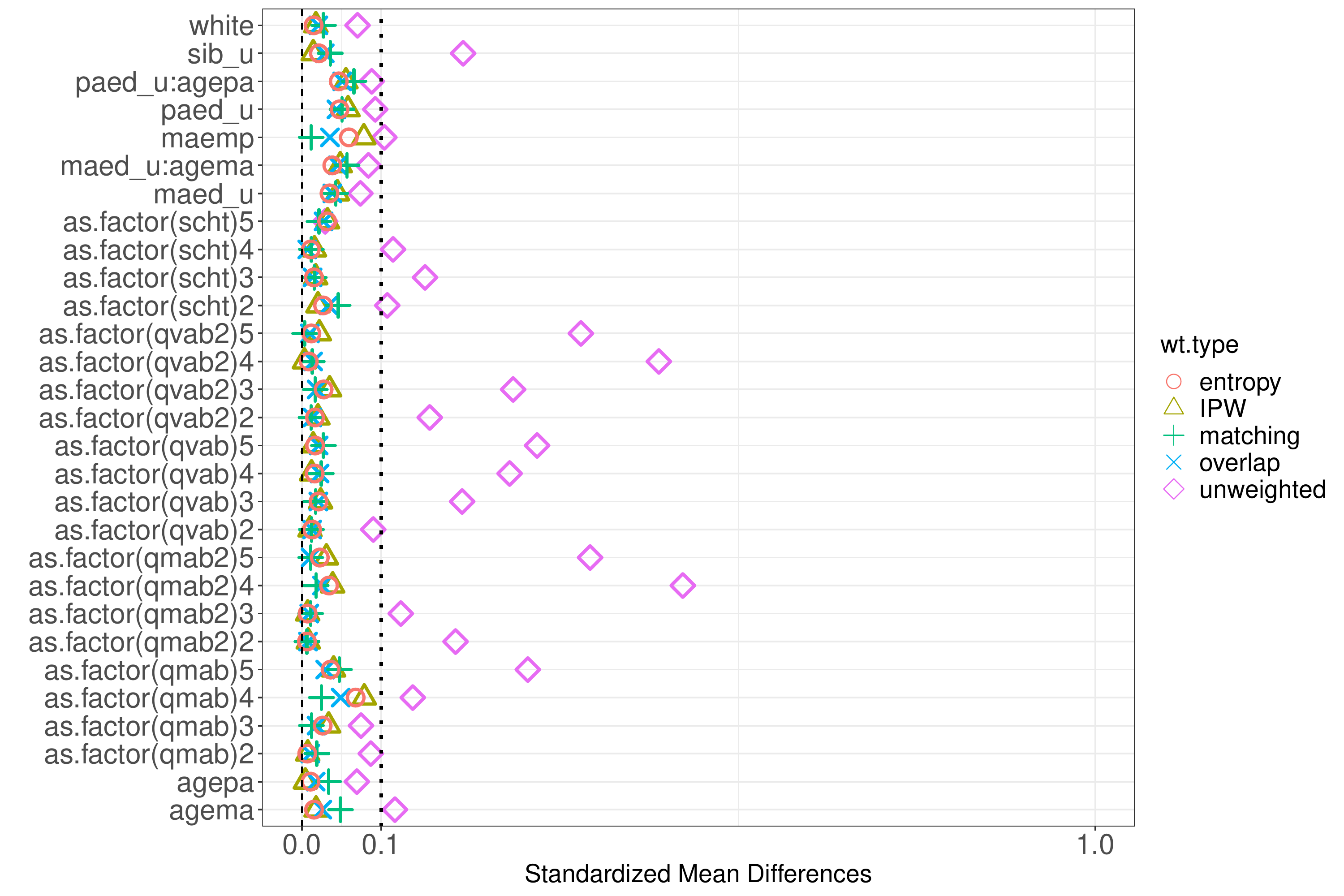}
\caption{Love plot with the three-level treatment variable \code{Dmult} using the maximum pairwise ASD metric, after symmetric trimming with $\delta=0.067$. This plot is generated by \code{plot.SumStat()} function in the \CRANpkg{PSweight} package.}\label{fig:mult4}
\end{figure}

The \CRANpkg{PSweight} package can perform trimming based on (generalized) propensity scores. As IPW does not adequately balance the covariates across the three groups in Figure \ref{fig:multasd}, we explore trimming as a way to improve balance for IPW. There are two types of trimming performed by the \CRANpkg{PSweight} package: (1) symmetric trimming that removes units with extreme (generalized propensity scores) \citep{Crump2009,Yoshida2019} and (2) optimal trimming that provides the most efficient IPW estimator for estimating (pairwise) ATE \citep{Crump2009,Yang2016}. Specifically, the symmetric trimming is supported by both the \code{SumStat()} and \code{PSweight()} functions through the \code{delta} argument. Both functions refit the (generalized) propensity score model after trimming following the recommendations in \citet{LiThomasLi2018}. We also provide a stand-alone \code{PStrim} function that performs both symmetric trimming and optimal trimming. Following \citet{Yoshida2019}, with three treatment groups, we exclude all individuals with the estimated generalized propensity scores less than $\delta=0.067$. This threshold removes a substantial amount of individuals in the advanced qualification group (information can be pulled from the \code{trim} element in the \code{SumStat} object). As discussed in \citet{Yoshida2019}, propensity trimming could improve the estimation of ATE and ATT, but barely have any effect for estimation of ATO and ATM. Evidently, Figure \ref{fig:mult4} indicates that IPW controls all pairwise ASD within $10\%$ in the trimmed sample. Trimming had nearly no effect on the weighted balance for OW, MW and EW.

\begin{Scode}
bal.mult.trim <- SumStat(ps.formula = ps.mult, weight = c("IPW", "overlap", "matching", 
    "entropy"), data = NCDS, delta = 0.067)

bal.mult.trim
\end{Scode}
 
\begin{Soutput}
1050 cases trimmed,  2592 cases remained 

trimmed result by trt group: 
         >=A/eq None O/eq
trimmed     778   71  201
remained   1028  824  740

weights estimated for:  IPW overlap matching entropy    
\end{Soutput}

\begin{Scode}
plot(bal.mult.trim,metric = "ASD")
\end{Scode}

Alternatively, if one does not specify the trimming threshold, the \code{PStrim} function supports the optimal trimming procedure that identifies the optimal threshold based on data. An example syntax is given as follows. By pulling out the summary statistics for trimming, we can see that optimal trimming excludes $27\%$, $9\%$ and $2\%$ of the individuals among those with advanced qualification, intermediate qualification and no qualification, respectively. The exclusion is more conservative compared to symmetric trimming with $\delta=0.067$. However, the resulting covariate balance after optimal trimming is similar to Figure \ref{fig:mult4} and omitted.

\begin{Scode}
PStrim(ps.formula = ps.mult, data = NCDS, optimal = TRUE)
\end{Scode}

\begin{Soutput} 
         >=A/eq None O/eq
trimmed     479   21   82
remained   1327  874  859
\end{Soutput} 

\subsubsection{Estimation and inference of pairwise (weighted) average treatment effects}

For illustration, we estimate the ratio estimands introduced in Section \ref{sec:ratioest} using the binary outcome \code{wagebin}. For illustration, we will only estimate the causal effects based on the data without trimming, and the analysis with the trimmed data follows the exact same steps. Based on the multinomial logistic propensity score model, we obtain the pairwise causal RR among the combined population via IPW.

\begin{Scode}
ate.mult <- PSweight(ps.formula = ps.mult, yname = "wagebin", data = NCDS,
    weight = "IPW")}
contrasts.mult <- rbind(c(1,-1, 0), c(1, 0,-1), c(0, -1, 1))
sum.ate.mult.rr <- summary(ate.mult, type = "RR", contrast = contrasts.mult)
sum.ate.mult.rr
\end{Scode}

\begin{Soutput}
Closed-form inference: 

Inference in log scale: 
Original group value:  >=A/eq, None, O/eq 

Contrast: 
          >=A/eq None O/eq
Contrast 1      1   -1    0
Contrast 2      1    0   -1
Contrast 3      0   -1    1

            Estimate Std.Error       lwr     upr  Pr(>|z|)    
Contrast 1  0.607027  0.115771  0.380120 0.83393 1.577e-07 ***
Contrast 2  0.459261  0.052294  0.356767 0.56176 < 2.2e-16 ***
Contrast 3  0.147766  0.121692 -0.090746 0.38628    0.2246    
---
Signif. codes:  0 ‘***’ 0.001 ‘**’ 0.01 ‘*’ 0.05 ‘.’ 0.1 ‘ ’ 1
\end{Soutput}

By providing the appropriate contrast matrix, we obtain all pairwise comparisons of the average potential outcomes on the log scale with the \code{summary.PSweight()} function, and estimate $\lambda^{h,\RR}(\pmb{a})$ for contrast vector $\pmb{a}$. The p-values provides statistical evidence against the weak causal null $H_0:\lambda^{h,\RR}(\pmb{a})=0$. It is found that, among the combined population, the proportion that receives above-average hourly wage when everyone attains advanced qualification is $\exp(0.607)=1.83$ times that when everyone attains no academic qualification. Further, the proportion that receives above-average hourly wage when everyone attains advanced qualification is $\exp(0.459)=1.58$ times that when everyone attains intermediate qualification. Both effects are significant at the $0.05$ levels and provides strong evidence against the corresponding causal null (p-value $<0.001$). However, if everyone attains intermediate qualification, the proportion that receives above-average hourly wage is only slightly higher compared to without qualification, with a p-value exceeding $0.05$. To directly report the causal RR and its confidence intervals, we can simply exponentiate the point estimate and confidence limits provided by the \code{summary.PSweight()} function. 

\begin{Scode}
exp(sum.ate.mult.rr$estimates[,c(1,4,5)])
\end{Scode}

\begin{Soutput}
          Estimate       lwr      upr
Contrast 1 1.834968 1.4624601 2.302358
Contrast 2 1.582904 1.4287028 1.753749
Contrast 3 1.159241 0.9132496 1.471493
\end{Soutput}

Focusing on the target population that has the most overlap in the observed covariates, we further use the OW to estimate the pairwise causal RR. OW theoretically provides the best internal validity for pairwise comparisons; Figure \ref{fig:mult4} also indicates that OW achieves better covariate balance among the overlap population. Exponentiating the results provided by the \code{summary.PSweight()} function, we observe each pairwise causal RR has a larger effect size among the overlap weighted population. Interestingly, among the overlap population, the proportion that receives above-average hourly wage when everyone attains intermediate qualification becomes approximately $1.55$ times that when everyone attains no academic qualification, and the associated $95\%$ CI excludes the null. Moreover, the standard errors for the pairwise comparisons are smaller when using OW versus IPW, implying that OW analysis generally corresponds to increased power by focusing on a population with equipoise. We repeat the analysis using both MW and EW; the results are similar to OW for this analysis and therefore omitted for brevity.

\begin{Scode}
ato.mult <- PSweight(ps.formula = ps.mult, yname = "wagebin", data = NCDS, 
    weight = "overlap")
sum.ato.mult.rr <- summary(ato.mult, type = "RR", contrast = contrasts.mult)
exp(sum.ato.mult.rr$estimates[,c(1,4,5)])
\end{Scode}

\begin{Soutput}
          Estimate      lwr      upr
Contrast 1 2.299609 1.947140 2.715882
Contrast 2 1.527931 1.363092 1.712705
Contrast 3 1.505048 1.257180 1.801785
\end{Soutput}

The above output suggests that among the overlap population, the causal RR for comparing advanced qualification and intermediate qualification is similar in magnitude to that for comparing intermediate qualification and no qualification. We can formally test for the equality of two consecutive causal RR based on the null hypothesis $H_0: \mu_3^h/\mu_2^h=\mu_2^h/\mu_1^h$ (also see Section \ref{sec:ratioest}). Operationally, we need to specify the corresponding contrast vector \code{contrast = c(1, 1, -2)}. The p-value for testing this null is $0.91$ (output omitted for brevity), and suggests a lack of evidence against the equality of consecutive causal RR at the $0.05$ level.

\begin{Scode}
summary(ato.mult, type = "RR", contrast = c(1, 1, -2), CI = FALSE)
\end{Scode}

With the binary outcome \code{wagebin}, we can also estimate the pairwise causal OR among a specific target population. For example, using OW, the causal conclusions regarding the effectiveness due to attaining academic qualification do not change, because all three $95\%$ confidence intervals exclude null. However, the pairwise causal OR appear larger than the pairwise causal RR. This is expected because our outcome of interest is not uncommon \citep{nurminen1995use}. For rare outcomes, causal OR approximates causal RR. 

\begin{Scode}
sum.ato.mult.or <- summary(ato.mult, type = "OR", contrast = contrasts.mult)
exp(sum.ato.mult.or$estimates[,c(1,4,5)])
\end{Scode}

\begin{Soutput}
          Estimate      lwr      upr
Contrast 1 3.586050 2.841383 4.525879
Contrast 2 2.050513 1.696916 2.477791
Contrast 3 1.748855 1.375483 2.223578
\end{Soutput}

As a final step, we illustrate how to combine OW with outcome regression and estimate the pairwise causal RR among the overlap population. 
We use the same set of covariates in the binary outcome regression model.

\begin{Scode}
 out.wagebin <- wagebin ~ white + maemp + as.factor(scht) + as.factor(qmab) +
    as.factor(qmab2) + as.factor(qvab) + as.factor(qvab2) + paed_u + maed_u +
    agepa + agema + sib_u + paed_u * agepa + maed_u * agema
\end{Scode}

Loading this outcome regression formula into the \code{PSweight()} function, and specifying \code{family = "binomial"} to indicate the type of outcome, we obtain the augmented overlap weighting estimates on the log RR scale. Exponentiating the point estimates and confidence limits, one reports the pairwise causal RR. The pairwise causal RR reported by the augmented OW estimator is similar to that reported by the simple OW estimator; further, the width of the confidence interval is also comparable before and after outcome augmentation, and the causal conclusions based on pairwise RR remain the same. The similarity between simple and augmented OW estimators implies that OW itself may already be efficient.

\begin{Scode}
ato.mult.aug <- PSweight(ps.formula = ps.mult, yname = "wagebin", data = NCDS,
    augmentation = TRUE, out.formula = out.wagebin, family = "binomial")
sum.ato.mult.aug.rr <- summary(ato.mult.aug, type = "RR", contrast = contrasts.mult)
exp(sum.ato.mult.aug.rr$estimates[,c(1,4,5)])
\end{Scode}

\begin{Soutput}
          Estimate      lwr      upr
Contrast 1 2.310628 1.957754 2.727105
Contrast 2 1.540176 1.375066 1.725111
Contrast 3 1.500237 1.253646 1.795331
\end{Soutput}

\subsection{Using machine learning to estimate propensity scores and potential outcomes}\label{sec:impillu}

As an alternative to the default generalized linear models, we can use more advanced machine learning models to estimate propensity scores and potential outcomes. Flexible propensity score and outcome estimation has been demonstrated to reduce bias due to model misspecification, and potentially improve covariate balance \citep{lee2010improving,hill2011bayesian,McCaffrey2013}. This can be achieved in \CRANpkg{PSweight} for both balance check and constructing weighted estimator by specifying the method as the generalized boosted model (GBM) or the super learner methods.  Additional model specifications for these methods can be supplied through \code{ps.control} and \code{out.control}. Machine learning models that are included in neither \code{gbm} nor \code{SuperLearner} could be estimated externally and then imported through the \code{ps.estimate} and \code{out.estimate} arguments. These two arguments broaden the utility of \CRANpkg{PSweight} where any externally generated estimates of propensity scores and potential outcomes models can be easily incorporated.

We now illustrate the use of GBM as an alternative of the default generalized linear models. For simplicity, this illustration is based on binary education. Specifically, we created \code{Dany} to indicate whether one had attained any academic qualification. There are 2,399 individuals that attained academic qualification, and 1,243 individuals without any. GBM is a family of non-parametric tree-based regressions that allow for flexible non-linear relationships between predictors and outcomes \citep{Friedman2000}. The following propensity model formula is specified; the formula does not include interactions terms because boosted regression is already capable of capturing non-linear effects and interactions \citep{McCaffrey2004}. In this illustration, we use the AdaBoost \citep{freund1997decision} algorithm to fit the propensity model through the control setting, \code{ps.control=list(distribution = "adaboost")}. We use the default values for other model parameters such as the number of trees (\code{n.trees = 100}), interaction depth (\code{interaction.depth = 1}), the minimum number of observations in the terminal nodes (\code{n.minobsinnode = 1}), shrinkage reduction (\code{shrinkage = 0.1}), and bagging fraction (\code{shrinkage = 0.5}). Alternative values for these parameters could also be passed through \code{ps.control}.  

\begin{Scode}
ps.any.gbm <- Dany ~ white + maemp + as.factor(scht) + as.factor(qmab) + 
    as.factor(qmab2) + as.factor(qvab) + as.factor(qvab2) + paed_u +  maed_u+ 
    agepa + agema + sib_u
bal.any.gbm <-SumStat(ps.formula = ps.any.gbm, data= NCDS,  weight = "overlap",
    method = "gbm", ps.control = list(distribution = "adaboost"))
\end{Scode}

The balance check through \code{plot.SumStat()} suggests substantial improvement in covariate balance with SMD of all covariates below 0.1 after weighting. After assessing balance and confirming the adequacy of the propensity score model, we further fit the outcome model using GBM with the default logistic regression and parameters. In the \code{PSweight()} function, we can specify both \code{ps.method = "gbm"} and \code{out.method = "gbm"} and leave the \code{out.control} argument as default. The detailed code and summary of the output is in below. Here we redefine the propensity score model without interaction terms because GBM considers interactions between covariates by default. The results using GBM, in this example, are very similar to those using generalized linear models (results omitted).

\begin{Scode}
out.wage.gbm <- wage ~ white + maemp + as.factor(scht) + as.factor(qmab) +
    as.factor(qmab2) + as.factor(qvab) + as.factor(qvab2) + paed_u + 
    maed_u + agepa + agema + sib_u
ato.any.aug.gbm <- PSweight(ps.formula = ps.any.gbm, yname = "wagebin",
    data = NCDS,  augmentation = TRUE, out.formula = out.wage.gbm, 
    ps.method = "gbm", ps.control = list(distribution = "adaboost"), 
    out.method = "gbm")
summary(ato.any.aug.gbm, CI = FALSE)
\end{Scode}

\begin{Soutput}
Closed-form inference: 

Original group value:  0, 1 

Contrast: 
            0 1
Contrast 1 -1 1

          Estimate Std.Error z value  Pr(>|z|)    
Contrast 1 0.186908  0.018609  10.044 < 2.2e-16 ***
---
Signif. codes:  0 ‘***’ 0.001 ‘**’ 0.01 ‘*’ 0.05 ‘.’ 0.1 ‘ ’ 1
\end{Soutput}

\section{Summary} \label{sec:summary}

Propensity score weighting is an important tool for causal inference and comparative effectiveness research. This paper introduces the \CRANpkg{PSweight} package and demonstrates its functionality with the NCDS data example in the context of binary and multiple treatment groups. In addition to providing easy-to-read balance statistics and plots to aid the design of observational studies, the \CRANpkg{PSweight} offers point and variance estimation with a variety of weighting schemes for the (weighted) average treatment effects on both the additive and ratio scales. These weighting schemes include the optimal overlap weights recently introduced in \citet{LiMorganZaslavsky2018} and \citet{li2019propensity}, and could help generate valid causal comparative effectiveness evidence among the population at equipoise. 

Although propensity score weighting has been largely developed in observational studies, it is also an important tool for covariate adjustment in randomized controlled trials (RCTs). \citet{williamson2014variance} showed that IPW can reduce the variance of the unadjusted difference-in-means treatment effect estimator in RCTs, and \citet{shen2014inverse} proved that the IPW estimator is semiparametric efficient and asymptotically equivalent to the analysis of covariance (ANCOVA) estimator \citep{tsiatis2008covariate}. \citet{zeng2020RCT} generalized these results of IPW to the family of balancing weights. Operationally, there is no difference in implementing propensity score weighting between RCTs and observational studies. Therefore, \CRANpkg{PSweight} is directly applicable to perform covariate-adjusted analysis in RCTs.

The \CRANpkg{PSweight} package is under continuing development to include other useful components for propensity score weighting analysis. Specifically, future versions of \CRANpkg{PSweight} will include components to enable pre-specified subgroup analysis with balancing weights and flexible variable selection tools \citep{yang2020propensity}. We are also studying overlap weighting estimators with time-to-event outcomes and complex survey designs. Those new features are being actively developed concurrently with our extensions to the methodology.

\section{Acknowledgement}
The authors would like to acknowledge the NCDS replication data published on Harvard Dataverse (\url{https://dataverse.harvard.edu/}) \citep{DVNEPCYUL2012}, which provides a coded data set for our analysis in Section \ref{sec:illustrations}.

\bibliography{PSweight}

\address{Tianhui Zhou\\
  Department of Biostatistics and Bioinformatics\\
  Duke University School of Medicine\\
  2424 Erwin Road, Suite 1105\\
  Durham, NC 27705, United States of America\\
  E-mail: \email{tianhui.zhou@duke.edu}}

\address{Guangyu Tong\\
  Department of Biostatistics\\
  Yale University School of Public Health\\
  300 George St, Suite 511\\
  New Haven, CT 06510, United States of America\\
  E-mail: \email{guangyu.tong@yale.edu}}
  
\address{Fan Li\\
  Department of Statistical Science\\ 
  Duke University\\ 
  122 Old Chemistry Building\\ 
  Durham, NC 27705, United States of America\\
  E-mail: \email{fl35@duke.edu}}

\address{Laine E. Thomas\\ 
  Department of Biostatistics and Bioinformatics\\
  Duke University School of Medicine\\
  2424 Erwin Road, Suite 1105\\
  Durham, NC 27705, United States of America\\
  E-mail: \email{laine.thomas@duke.edu}}

\address{Fan Li\\ 
  Department of Biostatistics\\
  Yale University School of Public Health\\
  135 College St, Suite 200\\
  New Haven, CT 06510, United States of America\\
  E-mail: \email{fan.f.li@yale.edu}}

\end{article}

\end{document}